# Low resistive edge contacts to CVD-grown graphene using a CMOS compatible metal

*Mehrdad Shaygan [1]\*, Martin Otto[1], Abhay A. Sagade [1,2], Carlos A. Chavarin[3,4], Gerd Bacher[3], Wolfgang Mertin[3], Daniel Neumaier[1]*

\*Corresponding Author: E-mail: shaygan@amo.de, m.shaygan@gmail.com

[1] Advanced Microelectronic Center Aachen, AMO GmbH, 52074 Aachen, Germany
[2] Center for Advanced Photonics and Electronics, Department of Engineering, 9 JJ Thomson Avenue, University of Cambridge, Cambridge CB3 0FA, UK.
[3] Werkstoffe der Elektrotechnik, Faculty of Engineering, University Duisburg-Essen, 47057 Duisburg, Germany
[4] Innovations for High Performance Microelectronics, IHP GmbH, 15236 Frankfurt (Oder), Germany

The exploitation of the excellent intrinsic electronic properties of graphene for device applications is hampered by a large contact resistance between the metal and graphene. The formation of edge contacts rather than top contacts is one of the most promising solutions for realizing low ohmic contacts. In this paper the fabrication and characterization of edge contacts to large area CVD-grown monolayer graphene by means of optical lithography using CMOS compatible metals, i.e. Nickel and Aluminum is reported. Extraction of the contact resistance by Transfer Line Method (TLM) as well as the direct measurement using Kelvin Probe Force Microscopy demonstrates a very low width specific contact resistance down to 130 $\Omega\mu m$. The contact resistance is found to be stable for annealing temperatures up to 150°C enabling further device processing. Using this contact scheme for edge contacts, a field effect transistor based on CVD graphene with a high transconductance of 0.63 mS/μm at 1 V bias voltage is fabricated.

# 1. Introduction

Graphene [1,2] has attracted scientific and technological interest that has been fueled by its unique properties such as high intrinsic mobility [3], high saturation velocity [4] and large breakdown current density [5]. Especially the potential integration of graphene-based electronic and photonic devices into a silicon complementary metal-oxide-semiconductor (CMOS) platform is very promising for future applications, as those hybrid systems can exploit the benefits of both materials: i.e. the well-developed fabrication platform for Si and the outstanding performance of graphene-based devices [6,7]. As a material basis for realizing such devices on Si, graphene grown by chemical vapor deposition (CVD) on a catalytic metal together with a subsequent transfer process [8] is currently one of the best options, as the direct growth of graphene on Si or $SiO_2$ does not yet provide sufficient material quality [9,10]. However, significant challenges related to the fabrication process limit the exploitation of the ultimate performance potential of graphene-based devices in a CMOS environment. One major issue, especially for graphene devices with sub-micrometer dimensions, is the contact resistance between the graphene and the contact metal [11]. So far, significant efforts have been made to reduce the contact resistance. For instance, the contact metal itself has a significant impact on the contact resistance because of the work-function difference and the bonding strength between the metal and graphene [12-16]. Also, impurities and contamination at the graphene-metal interface affect the contact resistance, and different cleaning and pretreatment procedures have been developed in order to minimize these effects [17-19]. Especially for graphene grown by CVD, surface contamination is a big problem because the transfer process to the target substrate already introduces a significant amount of polymer residuals [20]. An alternative route for realizing low ohmic contacts to CVD-grown graphene avoiding the impact of surface contamination is the formation of edge contacts [21], which has been demonstrated to provide a width specific contact resistance $R_cW$ in the range of 100 to 1900 $\Omega\mu m$ [21,22]. However, these relatively low $R_cW$ values have only been realized using noble metals like Au or Pd, which are not compatible with standard CMOS processing. In this paper we report on an efficient procedure for fabricating edge contacts to large area CVD-grown graphene with $R_cW$ down to 130 $\Omega\mu m$ using Ni/Al contacts, two metals widely used in CMOS processing. Using Transfer Length Method (TLM) [23] electrical characterization and in addition Kelvin Probe Force Microscopy (KPFM), we conducted a systematic study to extract $R_c$ at the metal-graphene (M-G) interface for top- and edge-contacted devices. While TLM measurements are widely used to extract $R_c$ in graphene devices, a large fitting uncertainty often does not allow a precise extraction of $R_c$. Additionally, by using TLM the measurement of $R_c$ of an individual metal-graphene contact is not possible. Therefore, we studied the voltage drop of individual two-terminal devices by means of KPFM for a more precise measurement of the contact resistance and to investigate the contact to contact variations of the contact resistance. Finally, the direct impact of the low contact resistance on the performance of field effect transistors is demonstrated by showing a high transconductance value in a graphene based field effect transistor (GFET) using edge contacts.

**2. Materials and Methods**

TLM structures for extracting the contact resistance were fabricated by means of optical contact lithography using AZ5214E as the resist. First, commercial graphene grown by CVD on copper foil (Graphenea SE) was transferred onto a 90 nm $SiO_2$ / highly doped Si substrate by wet-chemical etching of copper using $FeCl_3$ and polymethyl-methacrylate (PMMA) as support layer [24]. For the edge contact devices, the fabrication starts with defining the resist mask for the contacts and etching the graphene for 30s with oxygen plasma in a barrel reactor at 100 W as shown in Fig.1a. Metal contacts of Ni/Al (15/120 nm) were fabricated using sputter deposition followed by a lift-off process while keeping the same resist mask, which ensures the alignment. Nickel was chosen as the metal in contact to the graphene because simulations predict a very low contact resistance of nickel to graphene among metals compatible with CMOS technology [15]. According to first principle density functional theory the strong spin filtering is responsible for the smaller contact resistance of edge-contact compared to top-contacted graphene with Ni contact [25]. Finally, using another resist mask, the graphene channel was patterned with oxygen plasma. For comparison, reference samples with top-contacted graphene devices were fabricated in which the graphene channel was first patterned by oxygen plasma and then contacted with 15/120 nm of Ni/Al using sputtering deposition and lift-off. All electrical measurements were performed at room temperature under nitrogen atmosphere using a needle probe station and a semiconductor parameter analyzer (HP 4156B).

The KPFM analysis to directly measure the contact resistance was performed under ambient conditions in a NT-MDT NTEGRA Spectra system. Single pass phase modulated KPFM was used, applying an AC voltage of 2 $V_{p\text{-}p}$ with a frequency of 2.5 kHz to the tip, while the source contact of the device was grounded. The drain-source voltage ($V_{ds}$) was increased in 0.5 V steps up to ± 2 V. To obtain the voltage drop across the contacts and the graphene channel, the KPFM results at $V_{ds} = 0$ V were subtracted from that at $V_{ds} \neq 0$ V. After each measurement, a new KPFM map at $V_{ds} = 0$ V was made to reduce the influence of charging or aging effects. Silicon probes from Budgets sensors with a Cr/Pt/Ir coating and a nominal radius of < 25 nm were used.

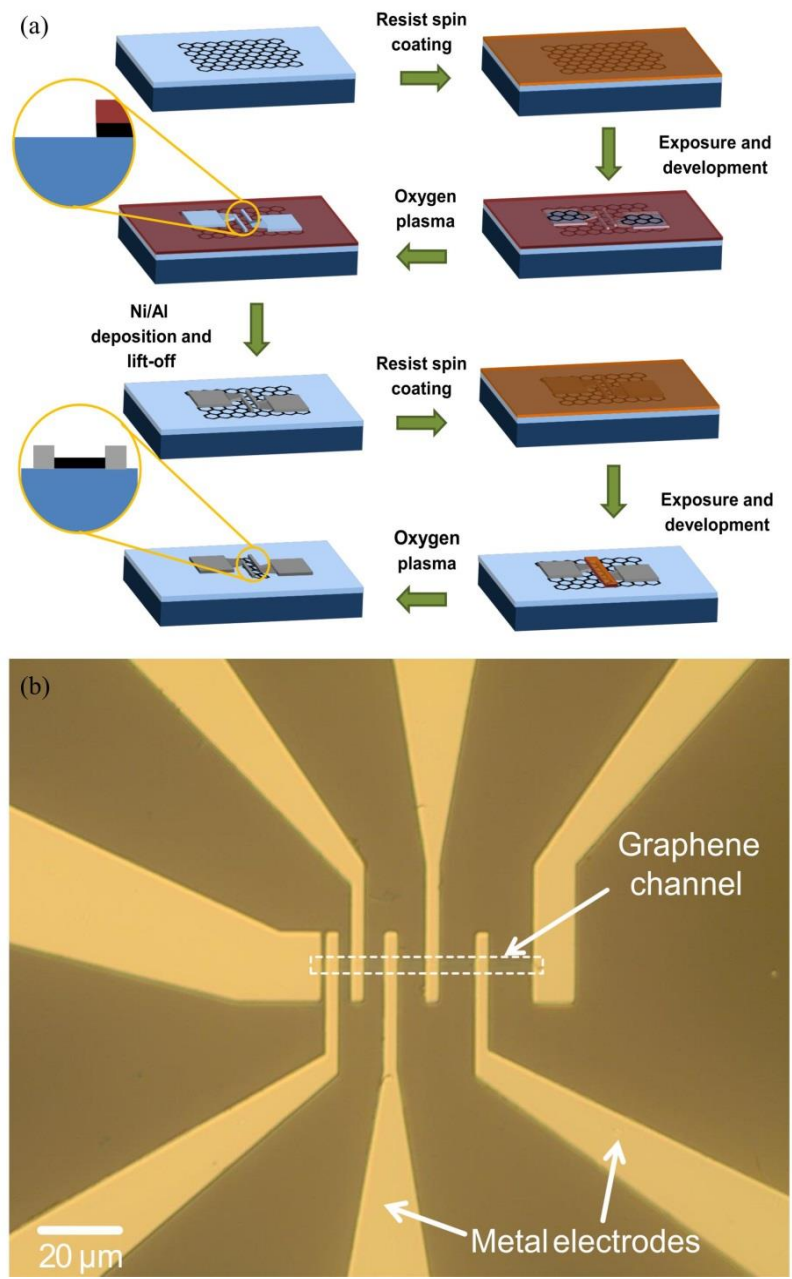

**Figure 1.** (a) Schematic view of the device fabrication procedure for realizing the edge contacts. (b) Optical micrograph of a fabricated graphene TLM structure on a SiO$_2$/Si substrate.

## 3. Results and Discussion

Figure 1b depicts an optical micrograph of the fabricated graphene TLM structure with contact spacing ($L$) ranging from 1.2 and 11.2 μm and a channel width ($W$) of 5 μm. Because of the limitations of optical contact lithography, the contact spacing and channel width varied slightly (up to 500 nm) from sample to sample. Therefore, for all samples the final geometry used for data analysis was measured after the electrical measurements with a scanning electron microscope.

The total resistance ($R_{total}$) of each neighboring contact pair was measured as a function of back-gate voltage ($V_{gs}$), with the $V_{ds}$ set to 20 mV. Fig 2a and 2c shows the total resistance normalized by the width as a function of $V_{gs}$ for top-contacted and edge-contacted devices, respectively, for different contact spacings. All devices show the typical behavior for graphene FETs with the resistance maximum indicating the charge neutrality point ($V_{CNP}$). Also, for both top- and edge-contacted devices, we observe a dependency of the $V_{CNP}$ on contact spacing, where longer spacing shows higher $p$-type doping. This channel length dependency of $V_{CNP}$ has already been observed before and can be related to contact-induced doping [26,27]. The total width specific resistance at $V_{CNP}$ and at $V_{gs}$ = -30 V is plotted as a function of channel length for top- and edge-contacted devices in Fig. 2b and 2d, respectively. In this plot the intersection of a linear fit with the y-axis gives two times $R_CW$, as $R_{total}W$ can be expressed by:

$$R_{total}W = (R_{sheet}L + 2R_cW), \qquad (1)$$

where $R_{sheet}$ is the sheet resistance of graphene. While for the top-contacted device we find a contact resistance of 2.5 ± 1 kΩμm at $V_{CNP}$ and 1.3 ± 0.3 kΩμm at $V_{gs}$ = -30 V, the contact resistance of the edge-contacted device is about 1 order of magnitude lower with a value of 360 ± 570 Ωμm at $V_{CNP}$ and less than 190 ± 120 Ωμm at $V_{gs}$ = -30 V. The errors here reflect the standard error of the linear fit. Overall we measured more than 10 TLM structures for top- and edge- contacted devices, fabricated in different runs and we always found a contact resistance of several kΩμm for top contacted devices, while for edge-contacted devices the contact resistance was always lower than 1 kΩμm, sometimes also giving negative values. While a striking difference could be identified in terms of contact resistance for top and edge-contacted devices using TLM measurements, the extraction of a reliable value of $R_c$ is not possible from the measurements of the TLM structures as can be seen from the large error of the linear fit, which is comparable to or even larger than the absolute value.

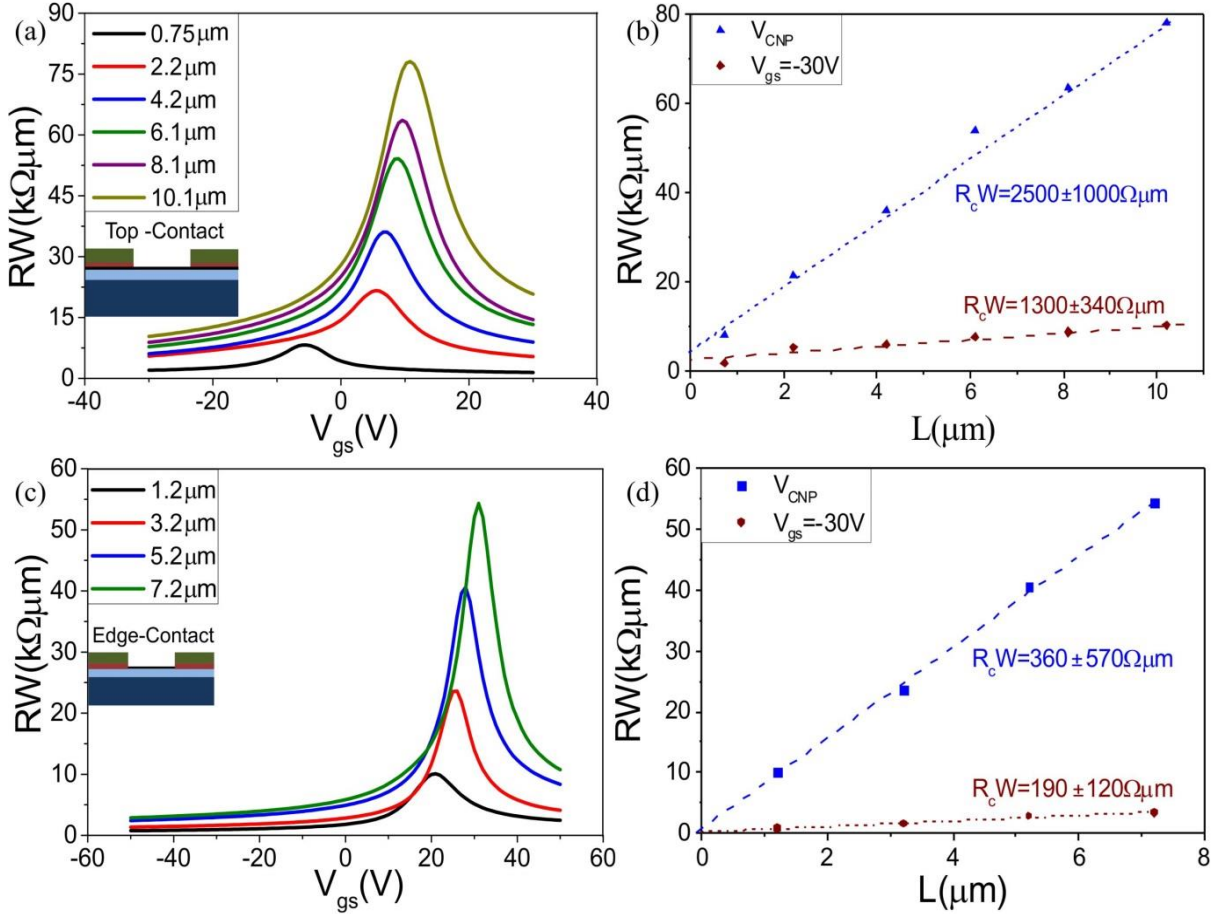

**Figure 2:** Electrical characterization of the TLM structures with top contacts (a,b) and edge contacts (c,d); (a,c) Width specific total resistance versus gate voltage, (b,d) Width specific total resistance for different channel lengths at $V_{gs} = -30\ V$ and at the $V_{CNP}$.

For extracting the contact resistance from TLM measurement in general, one needs to assume that the contact resistance is identical for each metal-graphene contact in the measured TLM structure and that the resistance of the graphene channel is constant between two contacts and also across the entire TLM structure. Especially for CVD-grown graphene these conditions are typically not met, because of inhomogeneities already present in the starting material. In addition, metal induced doping leads to a channel resistance where the resistivity of the channel depends on the specific distance to the contact.

In order to more precisely determine the exact value of the contact resistance, we performed KPFM voltage drop measurements on edge-contacted devices, which give a direct insight into the resistance distribution in individual two-terminal devices and enable us to see the resistance of each individual metal/graphene contact. Fig. 3a shows the quantitative voltage drop distribution image of a representative device ($L = 1.5$ µm, $W = 4$ µm) for $V_{ds} = -1$ V. The ends of both contacts are indicated by the white dashed lines. Due to the subtle change of the color contrast without any abrupt changes, i.e. no significant voltage drop, it is already visible in this plot that the contact resistance has to be very low compared to the resistance of the graphene channel.

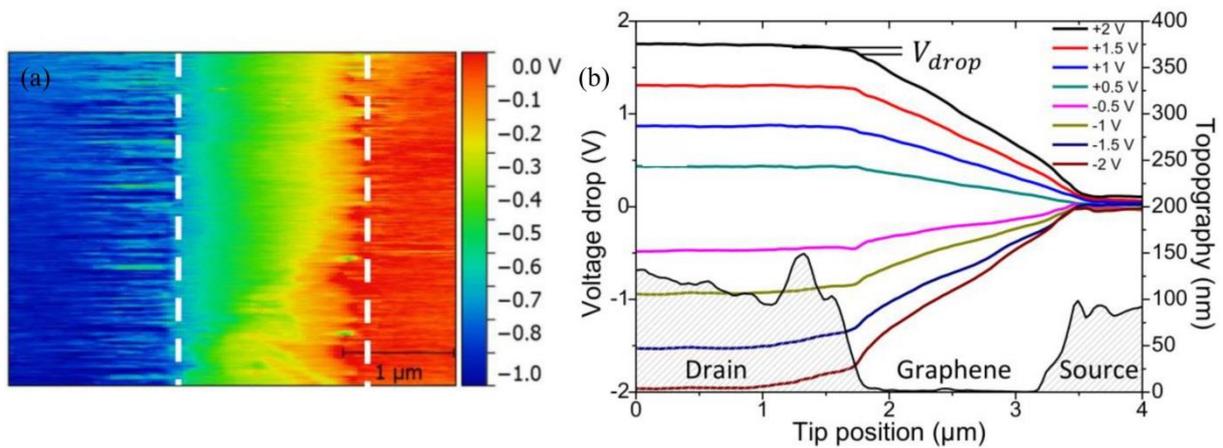

Figure 3. KPFM measurements. (a) Voltage drop image of an edge-contacted graphene device (W = 4 µm, L = 1.5 µm) for $V_{ds} = -1$ V, and (b) voltage drop profile vs. tip position at different drain-source voltages (averaged over 128 lines, the dashed area represents the corresponding topography of the sample). All measurements were done at $V_{gs} = 0$ V.

For a quantitative estimation, we extracted and averaged 128 lines (more than half of the channel's width) from the center of the map shown in Fig. 3a. The resulting voltage drop profile for $V_{ds} = -1$ V is plotted in Fig. 3b along with voltage drop profiles obtained from similar KPFM maps with bias voltages between -2 and 2 V. In this image, small and similar voltage drops are visible at the drain and source contacts. Note, the small voltage drop inside the drain contact is attributed to a screening effect related to a small residual layer nearby the metal edge as evidenced by the topography data. In order to quantify the potential drops across M-G interfaces and calculate the value $R_cW$, we used linear fit curves for the potential profiles in the contact and graphene regions and determined potential differences $V_{drop}$ at the edges of both contacts [28]. Using this simple model and the corresponding experimental value of the current $I_{ds}$ applied during the KPFM measurements, we were able to extract an $R_cW$ for each contact. For the device shown in Fig. 3, we extract a contact resistance of 280 ± 70 Ωμm and 380 ± 90 Ωμm for the left / drain and right / source contact, respectively. The error gives the standard deviation of the contact resistance extracted at 8 different bias voltages from -2 to 2 V (0.5 V steps). The values extracted by KPFM agree well with the values extracted by TLM, but they show a significantly higher accuracy. For comparison and as shown in our previous work top-contacted devices also produced by optical lithography had a contact resistance of several kΩμm in KPFM measurements. A detailed characterization with KPFM and AFM tools revealed a 3 - 4 nm-thick residual layer as the origin for the high contact resistance values in top-contacted devices produced by optical lithography [29]. In total, we extracted the contact resistance of 6 different devices using KPFM, giving a contact resistance for 12 different graphene-metal contacts (see Fig. 4a).

As can be seen, the contact resistance for all contacts measured is well below 1 kΩμm, with most of the values below 500 Ωμm and some even below 200 Ωμm, demonstrating the reproducible realization of competitive contact resistance values. The lowest value we found using KPFM was 130±10 Ωμm for the drain contact in device #5. We note that the large error for device number 1 is related to the geometry of this device, which was the device with the largest spacing between source and drain contacts (9.5μm). As all devices were measured at the identical bias voltages (i.e. -2V to 2V) the absolute voltage drop across the contact was lowest for the longest device, leading to a larger uncertainty for extracting the contact resistance.

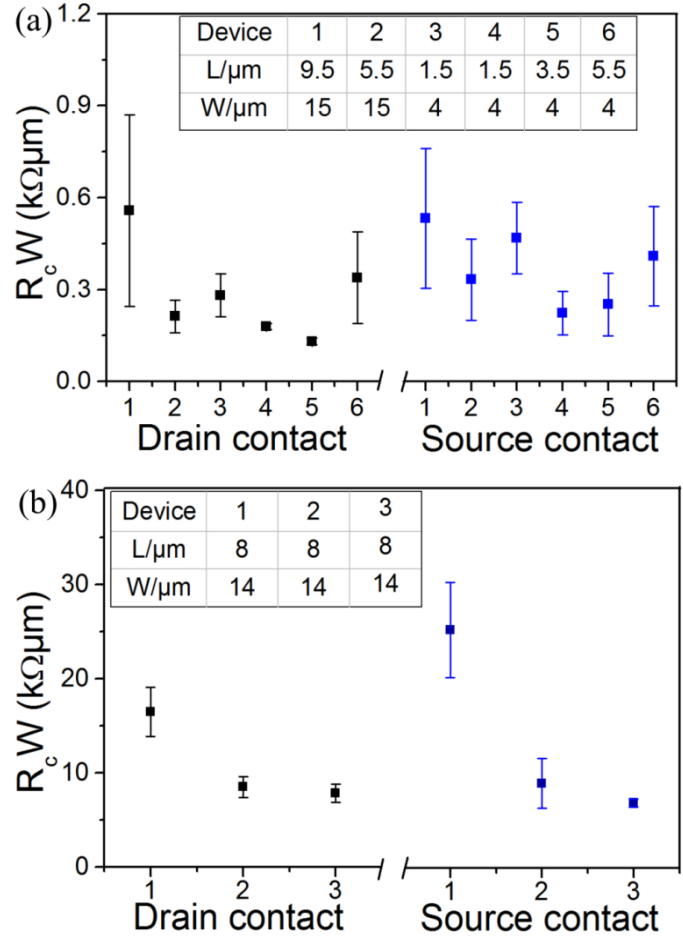

Figure 4. Width specific contact resistance extracted by KPFM in 6 devices for 12 contacts (a). The geometry of the different devices is given in the inset. The data points represent the average value and the error bar the standard deviation of the width specific contact resistance for one graphene-metal contact extracted at 8 different bias voltages from -2 to 2V. The device discussed in Fig. 3 is device #3 here. For comparison, corresponding width specific contact resistances for 3 devices with top contacts are shown in the panel (b) [Ref. 29].

Thermal stability of the contacts is mandatory for further device processing and during device operation. Therefore, one sample was annealed at 150 °C under Ar atmosphere for one hour and the contact resistance was measured before and after annealing using TLM measurements. 150°C was chosen because it is the highest temperature used in our process for graphene based integrated circuits after transferring the graphene layer to the sample [30,31]. Also, 150°C is above the typical operation temperature of electronic devices in end-user applications.

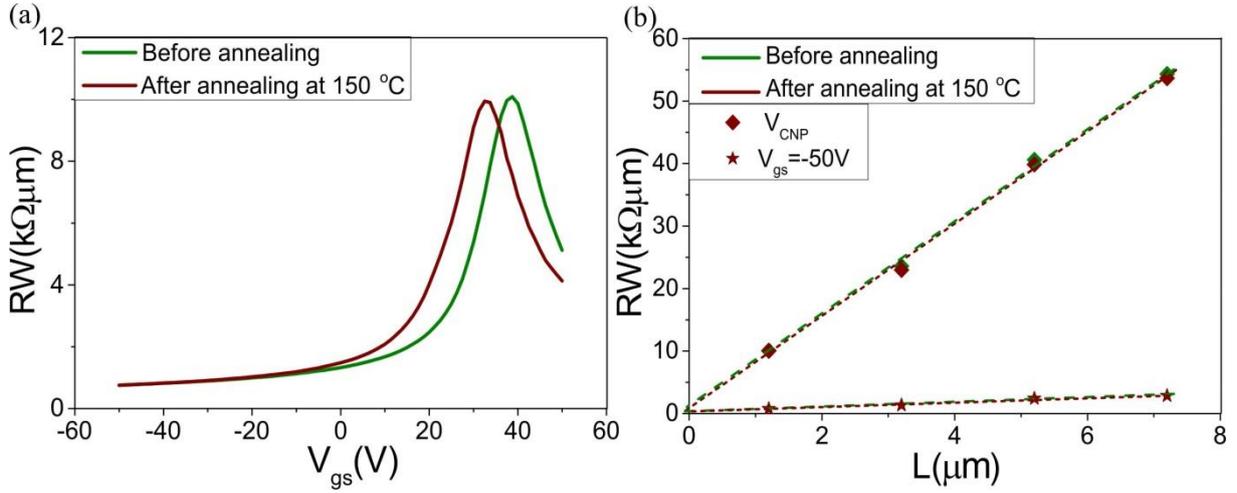

**Figure 5.** Effect of annealing on the electrical characteristic of the TLM structures with Ni/Al edge contacts. Total width specific resistance versus $V_{gs}$ of a 2µm long device (a) and total width specific resistance for different channel lengths at $V_{gs}$ = -50V and at $V_{CNP}$ (b).

Figure 5a shows the normalized resistance of one device with a contact spacing of 2 µm versus $V_{gs}$ before and after annealing. As can be seen, the doping slightly changes to lower *p*-type values, but the overall characteristic remains unchanged. The extraction of the contact resistance at $V_{gs}$ = -50 V and at $V_{CNP}$ shows that the contact resistance is unaffected by the annealing (Fig. 5b). $R_cW$ was 150 ± 100 Ωµm before and 150 ± 100 Ωµm after annealing measured at $V_{gs}$ = -50 V, and 350 ± 570 Ωµm and 300 ± 610 Ωµm, respectively, at the $V_{CNP}$, demonstrating that the edge contacts are stable at least up to temperatures of 150 °C.

To demonstrate the positive impact of a low contact resistance on the transconductance of GFETs, which is a figure of merit heavily affected by the contact resistance [32], we fabricated GFETs on Si substrates covered by 11 nm AlTiO as gate dielectric instead of 90 nm $SiO_2$. The dielectric was deposited by plasma-assisted atomic layer deposition at 300°C using trimethylaluminum (TMA), titanium chloride ($TiCl_4$) and oxygen as precursors. The dielectric constant of the AlTiO layer was 14, estimated by capacitance measurements on reference metal-oxide-metal structures using an LCR meter.

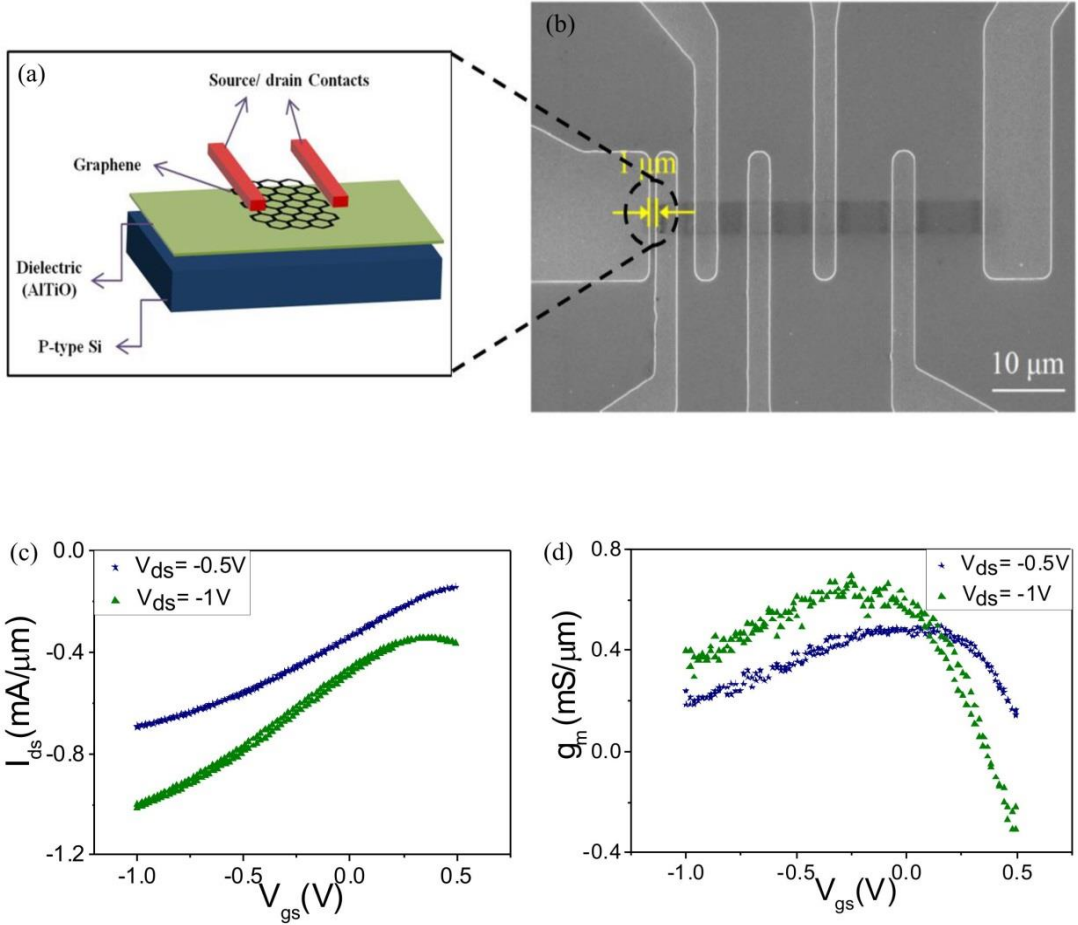

**Figure 6.** Schematic view (a) and SEM image (b) of the fabricated graphene FET. (c) Transfer characteristic and (d) transconductance at $V_{ds}$ of 0.5 and 1 V with a maximum of 0.63 mS/μm. Both sweep directions are plotted, demonstrating, that the hysteresis is negligible.

Figure 6a shows the schematic view of the fabricated graphene FET device. The device was fabricated with the above-mentioned technique for edge contacts. An SEM image of the device is depicted in Fig.6b. The transfer characteristic ($V_{ds}$ of -0.5 and -1 V) of the device with 1 μm contact spacing is depicted in Fig. 6c. The maximum transconductance of the device, $g_m = |dI_d/dV_g|$, at $V_{ds}$ = -1V showed a maximum value of 0.63mS/μm (Fig. 6d). Compared to devices with top contacts fabricated in our lab, where we did not achieve transconductance values higher than 0.25 mS/μm, this value is a significant improvement, resulting from the reduction of the contact resistance. In addition, a comparison with state-of-the-art transconductance values from literature is given in table 1, demonstrating that our device is competitive to the state-of-the-art for GFETs. We note that the transconductance is not only affected by the contact resistance, but depends particularly also on the equivalent oxide thickness, the mobility of the graphene, the applied voltage and the channel length.

**Table 1.** Comparison of maximum transconductance for different types of graphene FETs.

| FET Type | Max gm [mS/μm] | Vds [V] | Lg [μm] | Dielectric material | Dielectric Thickness [nm] | Ref |
|---|---|---|---|---|---|---|
| TG[a] epi[b] | 2 | 2.2 | 2.5 | $Si_3N_4$ | 15 | 33 |
| BG[c] CVD[d] | 1.2 | -1 | 0.5 | $HfO_2$ | 4 | 34 |
| BG CVD | 0.35 | -1 | 0.9 | $Al_2O_3$ | 10 | 35 |
| TG CVD | 0.53 | -0.6 | 0.3 | $Al_2O_3$ | N/A | 36 |
| BG CVD | 0.63 | -1 | 1 | AlTiO | 11 | This work |

[a] TG: top-gated FET; [b] epi: epitaxial graphene on SiC; [c] BG: back-gated FET; [d] CVD: CVD-grown graphene on metal;

## 4. Conclusion

In summary, we presented a procedure for making edge contacts to the graphene using a metal compatible with CMOS technology. Having contacted graphene with Ni/Al through optical lithography, a low width specific contact resistance of down to 130 Ωμm was achieved. This is a one order of magnitude reduction in contact resistance compared to top-contacted samples using the same metallization scheme. KPFM voltage drop measurements confirmed the values obtained by electrical TLM measurements. In addition, the reduction of $R_c$ leads to a significant improvement of the transconductance of GFETs with $g_m$ reaching 0.63 mS/μm.

**Acknowledgment**

The research leading to these results has received funding from the People Programme (Marie Curie Actions) of the European Union's Seventh Framework Programme FP7/2007-2013/

under REA grant agreement No. 607904-13 and the European Commission under the projects Graphene Flagship (contract No. 696656). D.N. and W.M. acknowledge funding from the German Research Foundation (DFG) in the frame of the SPP 1459 ''Graphene'' under contract NE 1633/2-1 and ME 1173/4-1. The work of C.A.C. was supported by The National Council on Science and Technology (CONACYT) Mexico under Grant No. 252826.

**Keywords**: Graphene field-effect transistor, contact resistance, edge contact, high-transconductance,